# Towards a More Accurate Carrier Sensing Model for CSMA Wireless Networks

Caihong Kai, Soung Chang Liew
Department of Information Engineering, The Chinese University of Hong Kong
Email: {chkai6, soung}@ie.cuhk.edu.hk

*Abstract*—This work calls into question a substantial body of past work on CSMA wireless networks. In the majority of studies on CSMA wireless networks, a contention graph is used to model the carrier sensing relationships (CS) among links. This is a "0-1" model in which two links can either sense each other completely or not. In real experiments, we observed that this is generally not the case: the CS relationship between the links are often probabilistic and can vary dynamically over time. This is the case even if the distance between the links is fixed and there is no drastic change in the environment. Furthermore, this "partial carrier sensing" relationship is prevalent and occurs over a wide range of distances between the links. This observation is not consistent with the 0-1 contention graph and implies that many results and conclusions drawn from previous theoretical studies need to be re-examined. This paper establishes a more accurate CS model with the objective of laying down a foundation for future theoretical studies that reflect reality. Towards that end, we set up detailed experiments to investigate the partial carrier sensing phenomenon. We discuss the implications and the use of our partial carrier sensing model in network analysis.

## I. INTRODUCTION

This paper concerns the carrier sensing behavior in wireless networks. Carrier sensing is an important feature of CSMA networks to avoid packet collisions. A node that has packets to send must first sense the channel. If no nearby node is transmitting, it transmits immediately. If a nearby node is transmitting, it defers, waiting until the end of the interfering transmission before attempting to transmit. That is, nodes that can sense each other will not transmit simultaneously in order that their packets do not collide.

With the widespread deployment of IEEE 802.11 networks, it is common today to find multiple wireless LANs co-located in the neighborhood of each other. The carrier sensing relationships among the links of these networks are **non-all-inclusive** in that not all the links can sense each other.

In the majority of prior studies, the carrier sensing relationships among the links are modeled by a contention graph. The links are represented by vertexes, and an edge joins two vertices if the associated links can sense each other. In other words, the carrier-sensing (CS) between two links is a 0-1 relationship in that they can either sense or not sense each other. A simple radio propagation model, two-ray ground model is often used, such that the CS relationship is a simple function of distance. If the distance between two links is shorter than a threshold (called the Carrier Sensing Range, CSRange), they can always sense each other; otherwise the two links will never hear each other. Although widely used in both theoretical and simulation studies, this model has not undergone rigorous verification in practice.

Indeed, our real-network experimental data showed that this "0-1" CS model is not accurate. In reality, two links can often only sense each other partially – they sometimes can hear each other and sometimes cannot.

To build a more accurate CS model, we have conducted extensive experiments to characterize the partial carrier sensing relationship among links. We show that a probabilistic CS model matches experimental data more closely than the prior absolute 0-1 CS model.

The main contributions of this paper are as follows: 1) we show that there is a long range of distance between two links over which the carrier sensing between them is partial rather than full, and therefore one can expect partial CS to be prevalent in a typical CSMA wireless networks; 2) we show that partial CS has a significant effect on link throughputs, and therefore one should not simply approximate the partial CS model with the 0-1 CS model in analysis; 3) we propose an accurate probabilistic partial CS model that match the experiment results; 4) we discuss implications and the use of the partial CS model in analytical studies.

*Related work*

Carrier sensing plays an important role in determining link throughputs in CSMA networks. Recent work, including [1-2], considered the tradeoff between spatial reuse and interference. Ref. [3,4] dealt with optimal choices of carrier sensing parameters for Gaussian signals. Ref. [5] studied the impacts of physical carrier sensing on system performance under slow fading channel where the packet collisions due to imperfect CS are considered. Although there are attempts to develop practical models for packet reception and carrier sensing (e.g., [6, 7]), the details of the carrier sensing mechanism have been ignored and only energy detection is taken into account.

In this paper we show that in reality the carrier sensing relationship is probabilistic over a long range of distance. Due to significant effects of partial carrier sensing on network performance, many previous theoretical and simulation frameworks need to be revisited. Some examples are as follows: 1) The NS2 simulator [8] is by far the most popular simulation tool used for the studies of 802.11 networks; however, it uses the unrealistic 0-1 carrier sensing model. 2) With partial carrier sensing, the throughput distributions among links in a CSMA network may be quite different from those derived under a 0-1 model. In particular, many reported problems, such as link starvation and unfairness [9] may be alleviated under partial carrier sensing, i.e., in practice things may not be as bad as predicted theoretically. 3) Many previous studies based on the 0-1 contention graph and their conclusions will need to be re-examined given the existence of partial carrier sensing. For example, the "island states" and the "phase transition phenomenon" as reported in [9] and [10] may not be common in practice.

The remainder of this paper is organized as follows. Section II explains that there is a significant transition range between full carrier sensing and no carrier sensing over which partial carrier sensing occurs. Section III examines our experimental results in detail and argues that partial carrier sensing is indeed the dominating factor behind our experimental observations. Section IV attempts to build an accurate carrier-sensing model based on experimental data of inter-packet arrival time. Section V discusses the implications and applications of partial carrier sensing. Section VI concludes this paper.

## II. LARGE TRANSITION RANGE OF PHYSICAL CARRIER SENSING IN REAL ENVIRONMENT

This section first gives a quick review of the carrier sensing mechanisms as defined in the IEEE 802.11 standard. After that, we show that there is a long range of distance between two links over which partial carrier sensing occurs. In particular, the transition from full carrier sensing to no carrier sensing is a gradual rather than an abrupt process as the distance varies.

### A. Carrier sensing in IEEE 802.11 standards

In 802.11 networks [11], physical carrier sensing (PCS) is performed by the Clear Channel Assessment (CCA) function, which monitors the channel to determine whether it is free. The 802.11 standard defines three CCA operation modes. The channel is declared as busy when (i) the energy detected exceeds a threshold $CS_{th}$; (ii) a valid 802.11 signal is detected, even if the power is below $CS_{th}$; (iii) either (i) or (ii) occurs.

The research community has largely considered only (i) in the investigations of CSMA networks, although modes (ii) and (iii) are often used in real 802.11 equipments.

Under (ii), in the event that a correct PLCP (Physical Layer convergence Procedure) Header is received, the CCA signal may be held inactive (channel busy) for the full duration of the packet as indicated by the PLCP LENGTH field. Even if a loss of carrier occurs in the middle of reception, the CCA will indicate a busy medium for the intended duration of the transmitted packet.

### B. Long transition range of carrier sensing

Much of the existing work assumes the two-ray ground model for the analysis of PCS relationships. Given a $CS_{th}$, the carrier sensing range (CSRange) is defined as the minimum distance between two transmitters such that concurrent transmissions are allowed. Within CSRange (550m in NS2 default setting), two links hear each other with probability 1; beyond CSRange, they can transmit independently.

Let us consider the normalized aggregate throughput of two short links as the link separation $d$ varies. Based on the 0-1 CS model, when $d <$ CSRange, the two links will share the channel and each of them will get half the medium airtime. The normalized aggregate throughput is thus 1. Once $d \geq$ CSRange, the two links can transmit as if they are isolated links. The normalized aggregate throughput jumps to 2 immediately. In this theoretical model, there is an abrupt jump at the critical point $d =$ CSRange (as shown on the left of Fig.1).

In real environment, we find there is a large, gradual transition range of the normalized aggregated throughput as $d$ increases. As shown on the right of Fig.1, the normalized aggregate throuhout of two links increases gradually from 1 to 2 as $d$ varies.

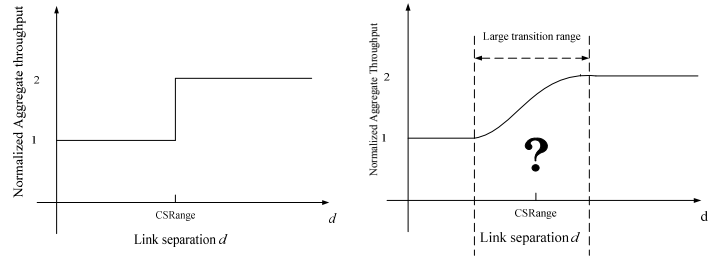

Fig.1. Long transition range between full carrier sensing and no carrier sensing.

Fig.2 shows the measured throughput of one of the links in a real 802.11a two-link network. Instead of an abrupt jump, the measured throughput increases gradually with $d$. We are interested in the underlying causes of this long transition range. One possibility is signal capture, which has been reported in [12]. Another possibility is partial carrier sensing, which we find to be the dominating factor over a long transition range, as will be explained in Section III.

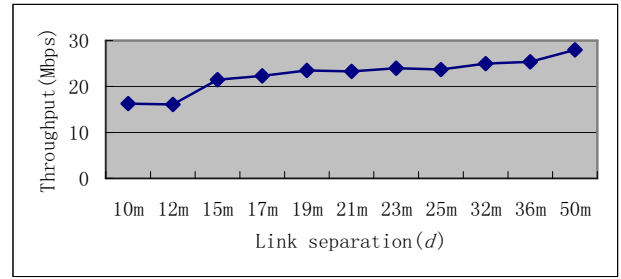

Fig.2. Measured throughput vs link separation $d$

## III. EXPERIMENT SETUP AND RESULTS

Our experiments are based on wireless cards that use the Atheros 802.11 chips. We want to measure the CS relationship between two links. However, we do not have direct access to the CCA information. We therefore design an indirect method for measuring CS. By looking at the variations of link throughputs, Packet Loss Ratios, numbers of transmission attempts per second according to different link separation $d$, we demonstrate the existence of partial carrier sensing.

### A. Experiment setup

We set up experiments with two pairs of DELL Latitude D505 laptops with 1.5GHz Celeron Mobile CPU. Each node has a NETGEAR WAG511v2 wireless card, and runs Fedora5 with MADWifi driver [13]. All Atheros chipset extensions were disabled. The network setup is shown in Fig.3. The distance between each sender-receiver pair was set to 0.1m to remove hidden-node effects. To make the experiment easier to control, the transmission power of each link was set to the minimum value allowed by hardware (1mW). Our experiments were conducted outdoor on 802.11a channel 36. OminiPeek, a network analysis software [14], was installed in another laptop to serve as a "sniffer" to collect traffic traces.

Typical 802.11a parameters were used in the experiments: (i) fixed data rate and basic rate of 54 Mbps and 6 Mbps, respectively; (ii) packet payload of 1460 Bytes; (iii) $CW_{min}$ of 15 and mini-timeslot of $9\mu s$, where CW is the contention window; (iv) basic mode of DCF. The transport protocol is UDP. Iperf, a network testing tool [15], was used to create UDP data streams and measure the throughputs. For each UDP session, the date rate was set to 30 Mbps to ensure link saturation. Each experiment lasted for 60 seconds and was repeated three times.

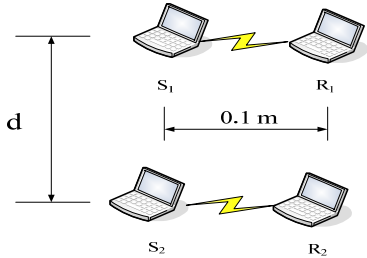

Fig.3. Experiment setup

### B. Experiment results

We define Packet Loss Ratio (PLR) as the ratio between the number of packets lost during transmission and the number of packets transmitted at the sender. That is, PLR = (# of packets transmitted at the sender - # of packets received at the receiver)/ # of packets transmitted at the sender. The retransmissions of the same packet at MAC layer is counted as multiple transmissions in our measurements. Table 1 lists the statistics of one of the links versus link separation $d$.

Table 1. Throughput & PLR vs link separation $d$

| Distance | 0.2m | 1m | 3m | 6m | 12m | 15m | 20m | 50m |
|---|---|---|---|---|---|---|---|---|
| Throughput (Mbps) | 14.6 | 15.6 | 16.11 | 16.22 | 16.3 | 22.7 | 23.7 | 27.8 |
| PLR | 12.9% | 5.0% | 1.0% | 1.0% | 0.7% | 1.5% | 1.5% | 0.4% |
| # of transmit attempts | 1412 | 1410 | 1393 | 1403 | 1404 | 1968 | 2055 | 2392 |

### C. Signal capture vs Partial carrier sensing

Physical-layer signal capture in 802.11 networks refers to the successful reception of the stronger (higher signal strength at receiver) packet in a collision [12]. For the experiment setup in Fig.3, when $d_{S_2 R_1} \gg d_{S_1 R_1}$, the power from $S_1$ to $R_1$ is much larger than that from $S_2$; hence, the transmitted packets of link 1 can be captured with good probability. In 802.11a networks the countdown time is uniformly chosen from [0, 15] timeslots, even if the CS is full, with probability $1/8.5 = 12\%$ two links can count down to zero and begin transmission in the same timeslot. When this happens, collisions may or may not happen depending on whether signal capture is in effect. When $d$ = 0.2m, the measured PLR is 12.9%. That is, there is no signal capture here. As $d$ increases, the measured PLR decreases, indicting that more packets can be captured due to stronger power. This, however, results in only slightly higher throughput.

1) Capture effect dominating range ($d$ = 0.2m ~ 12m)

As shown in Table 1, as $d$ increases from 0.2m to 12m, the number of transmission attempts does not vary much, indicating that CS is "full". In particular, the throughput increase does not come from increased transmission attempts; rather it is due to the smaller PLR as $d$ increases. We refer to this range of $d$ as the "capture effect dominating range". The throughput increase is smaller over this range.

2) Partial carrier sense dominating range ($d$ =12 ~ 50m)

As shown in Table 1, as $d$ increases from 12m to 50m, the throughput increase is much higher. PLR over this range remains more or less constant. The throughput increase mainly comes from increased number of transmission attempts. The ratio between throughput and number of transmission attempts is almost constant here. Since the transmitters can make more transmission attempts only when they cannot fully hear each other, we conclude that in this range partial carrier sensing kicks in. Also, as $d$ increases, when the two links simultaneously transmit, signal captures have a good chance to occur. However, signal capture alone cannot explain the large increase in throughput without partial carrier sensing.

The experiment data in Table 1 indicate that there is a large "partial carrier sense dominating range" over which the likelihood of carrier sensing varies from 100% to 0%, as explained below.

For an isolated link, the time consumed by a successful packet transmission consists of (i) PACKET duration consisting of physical-layer preamble/header, MAC Header, and data payload; (ii) SIFS; (iii) ACK; (iv) DIFS; (v) the random number of backoff countdown timeslots. For each packet, the airtime within its carrier-sensing range that must be exclusively dedicated to it is

$T_{tr}$ = PACKET + SIFS + ACK+DIFS     (1)

In addition, it also consumes a random backoff countdown time (i.e., component (v) above). Theoretically, in our setup the unshared time needed to transmit a packet is $340 \mu s$. So the theoretical throughput of an isolated link is $1460*8/(340+7.5*9) = 28.66$Mbps. If two links perfectly hear each other, the average time cycle needed to transmit a packet is the sum of the time consumed by two packets transmission plus a random backoff countdown time (Note that countdown time is shared by two links). Hence, the throughput of each link is $1460*8/(2*340+7.5*9) =15.63$ Mbps. With perfect signal capture, the throughput of each link should be $1460*8/((1+7.5/8.5)*340+7.5*9) =16.51$ Mbps.

The range of throughput between 16.51Mbps and 28.66Mbps as computed above map roughly to the range of throughputs between $d$ = 12m and $d$ = 50m, the partial carrier sense dominating range. The slightly lower experimental throughputs in Table 1 are attributed to the fact that our theoretical computation above ignores the periodic beacons sent out by APs and the random packet loss due to noise.

## IV. PARTIAL CARRIER SENSING MODELING

This section attempts to build an accurate carrier sensing model based on experimentally measured results.

### A. Measured inter-packet arrival time

As mentioned earlier, it is not easy to gather direct information on CCA from commercial wireless cards. In our experiments here, we collect the inter-packet arrival times at a sniffer using Omnipeek to capture the behavior of partial carrier sensing indirectly. Also, the retransmitted packets are regarded as successive arrival packets at the MAC layer in our measurements.

Since all the packets in our experiments are of the same length and are transmitted using the same data rate, the inter-packet arrival time at the sniffer is equal to the inter-packet transmission time at the transmitter plus measurement error.

(i) An isolated link

For an isolated link, the time needed to transmit a packet is 340 $\mu s$ or 38 timeslots ($340/9 \approx 38$). The backoff countdown time is uniformly distributed over [0, 15] timeslots. Hence, the inter-packet arrival time is uniformly distributed over [340, 475] $\mu s$ (i.e., over 340 $\mu s$ + [0, 15] random timeslots).

Fig. 4 plots the probability distributions of countdown time according to analysis and according to experiment. As shown, the experimental measurements match well with analysis with only very small deviations from the uniform distribution.

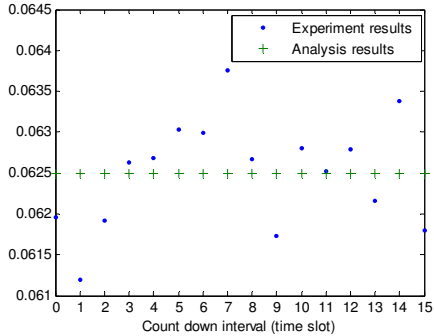

Fig.4. Countdown time distributions of an isolated link

(ii) Two links with full carrier sensing

When two links fully sense each other, once a link is frozen, it will be frozen for the whole packet transmission time. As a result, the inter-packet arrival time for packets of one link will fall into several bands: 340 $\mu s$ + [0, 15] timeslots (no freezing between two successive transmissions of the link); or 340 $\mu s$ + [38, 53] timeslots (frozen once between two transmissions); or 340 $\mu s$ + [76, 91] timeslots (frozen twice between two transmissions), and so on. In our real-network experiments shown in Section III, when $0m \leq d \leq 12m$, two links can fully sense each other. The countdown times (including the active and frozen countdown times) as measured from the inter-arrival times at the sniffer, fall into several bands. To verify the accuracy of our measurements, we compare the measured distribution of countdown time with that of theoretical computation in Fig. 5. As can be seen, the accuracy of our measurements is quite high.

(iii) Two links with partial carrier sensing

As argued in Section III, CS is partial when $12m < d < 50m$. We conducted extensive measurements on the inter-packet arrival times of one of the two links for different link separation $d$. Fig. 6 shows the distributions of countdown time, which we found to be quite stable over different measurement runs. As $d$ increases, more and more packets fall into the first band [0, 15]. After $d > 16m$, all the packets fall into a much wider first band of [0, 38] timeslots, indicating that the link never freezes for a whole packet's transmission time. Note from the difference between Fig. 4 (the isolated link case) and Fig. 6(d) that the link in Fig. 6(d) did get affected by the other link and did freeze from time to time (just that never more than one packet duration). In Section IV-B, we will attempt to build a more accurate model from the data presented in Fig. 6.

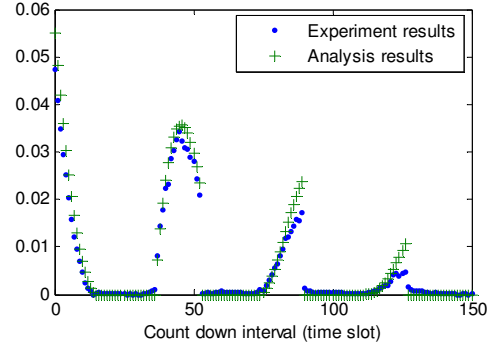

Fig.5. Countdown time distribution when full carrier sensing

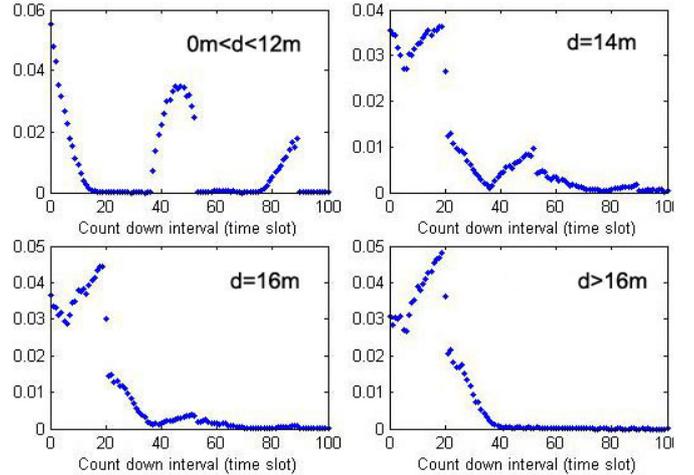

Fig.6. Countdown time distributions when partial carrier sensing

### B. A more accurate carrier sensing model

We find that a partial carrier sensing model as described in the next paragraph can match the experimental results in Fig. 6 rather well. In addition, the model is also compatible with the 802.11 standard specifications.

Over each timeslot, a node in idle state tries to detect the presence of a physical preamble with CS threshold set to -82dBm (receiver sensitivity required for data rate of 6 Mbps in 11a networks). Once a detection event is triggered, CCA will determine whether it is the start of a new 802.11 signal. If not, the node only freezes for this timeslot and continues sensing after that. Otherwise, it will spend time trying to track the carrier. It will do so for at least for 4 or 5 timeslots (time needed to receive a complete PHY header). If the PHY header is decoded successfully, the node will reserve the channel for the whole frame transmission time. If it can not decode the PHY header, then it adopts energy detection, with a threshold

20 dB above the minimum 6 Mbps sensitivity. Fig.7 shows the procedure of physical carrier sensing.

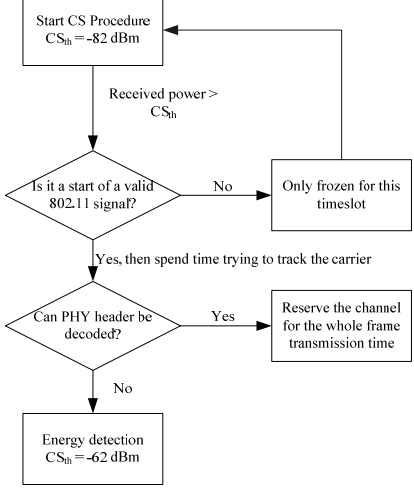

Fig. 7 The procedure of physical carrier sensing

In our special case of two links, when the transmitter of link 1 finishes a transmission and begins to count down, link 2 is either transmitting or counting down. If link 2 has been transmitting, obviously link 1 will not detect its PHY header because part of the packet of link 2 has already been transmitted. In this case, link 1 will keep sensing with CS threshold equal to -82dBm. Link 1 will be frozen with probability $p'$ in each of the subsequent timeslots.

If link 2 is also counting down, over each timeslot with probability $q$ link 2 begins to transmit first and link 1 detects this start of the physical preamble of link 2. Assuming the new generated countdown time of link 1 is $k$, which is uniformly distributed over [0, 15], then with probability $1-(1-q)^k$, link 1 will detect the transmission of link 2. Then link 1 spends 4-5 timeslots trying to track the carrier of link 2. We have the following two possibilities:

i) With probability $r$, the PHY header of link 2 can be decoded successfully. Once the PHY header is decoded successfully, link 1 will reserve the channel for the whole frame transmission time and its countdown will be frozen.

ii) With probability $1-r$, link 1 can not decode the PHY header of link 2. After that link 1 adopts energy detection, with a threshold 20 dB above the minimum 6 Mbps sensitivity. Since even the PHY header transmitted at 6 Mbps can not be decoded, when using a 20dB higher threshold, it is rare to find a busy channel to be frozen due to energy detection.

Fig.8 shows the comparison between experimental results and countdown time distributions under the partial CS model above when $d = 26m$. Our model fits the experimental curve very well when the parameters are set as follows: $r$ =0, $p'$ =0.47, $q$=0.04. Note and recall that when $d > 16m$, no packet falls into the second band. The link never reserves the channel for the whole frame transmission (i.e., $r$=0).

The CS model identified above is rather complex. It will be desirable to identify a simpler model that is amenable to analysis but which also captures the essence of partial CS.

From the model established above, we have the following observations:

*Observation 1:* The carrier sensing relationship changes *quickly* between "0" and "1". The sojourn time on either "0" or "1" is much shorter than a packet transmission time in most cases (when $d > 16m$);

*Observation 2:* With respect to the transmitter of one link, assuming the other link is transmitting, with some probability (1-$p$), it can also actively count down or begin transmitting.

We could build a simpler model where link $i$ can hear link $j$ with probability $p$ when link $j$ is transmitting. It is easy to see that $p$ decreases with the increase of link separation $d$.

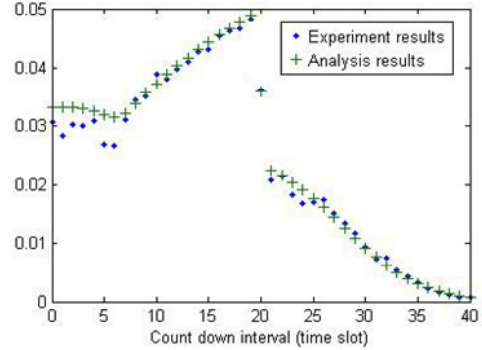

Fig.8. Comparison between experimental results and carrier sense model

## V. IMPLICATIONS AND APPLICATIONS OF PARTIAL CARRIER SENSING

Let us now consider the implications and applications of the partial carrier sensing model. We note that NS2 is arguably one of the most popular simulators for research on CSMA networks. In NS2, if two nodes are within CSRange, they can hear each other with probability 1; otherwise they cannot hear each other. Our partial carrier sensing results indicate the NS2 model is highly inaccurate in reality, and therefore results and conclusions drawn from the simulator are suspect. It would be desirable to modify NS2 for a more accurate CS model.

In the following we give three concrete applications to show the necessity of partial carrier sensing modeling.

### A. Throughput analysis

It is known that link throughput distributions of CSMA networks are quite unfair and extreme under the "0-1" CS modeling [9]. Starvations are prevalent. A link suffers from starvation, for example, when it is sandwiched between other links that keep transmitting. Consider the contention graph on the left of Fig.9. In this network, link 1 and link 3, link 1 and link 4 can transmit together. Together, either link 1 and link 3 or link 1 and link 4 grab the access to the channel most of the time, leaving little chance for link 2 to actively count down and transmit. This results in a normalized throughput distributions of [1, 0, 0.5, 0.5], where link 2 is starved [9]. However, if we assume each pair only can hear each other with probability $p = 0.8$, it can be shown that the normalized throughputs are [0.80, 0.25, 0.57, 0.57] (this computation is not presented here due to limited space). Link 2 will not be starved any more and better fairness can be achieved.

In a practical wireless network, due to partial carrier sensing, the starved links under the 0-1 model may actually obtain some throughputs. The unfairness and starvation problem may not be as bad as commonly reported in the literature.

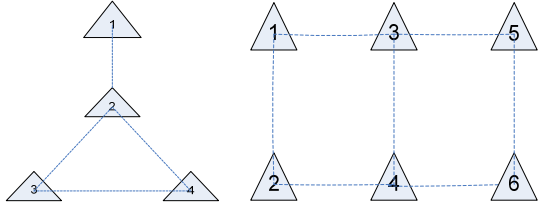

Fig.9. Two example networks

*B. Temporal starvation*

The "island state" and phase transition phenomena are observed when the 0-1 contention graph is used in analysis [9, 10]. In [9], it was shown that the CSMA system will spend most of its time in MIS (Maximum Independent States) and that each MIS is equally probable. Consider the topology on the right of Fig.9, the MIS are $\begin{smallmatrix}1&0&1\\0&1&0\end{smallmatrix}$ and $\begin{smallmatrix}0&1&0\\1&0&1\end{smallmatrix}$. The system will be in each state about half of the time, causing the long-term normalized throughput of each link to be 0.5. However, it has also been argued that once the system enters one MIS, it is difficult for it to transit to another state [9]. Temporal equilibrium stays around an MIS with only occasional movement across the two MIS. In this topology, three links can starve for a long time once the system settles around the MIS that disfavors them. All links suffer from temporal starvation here. This phenomenon may disappear in reality due to partial carrier sense. Consider a network of $L$ links. Because of partial carrier sensing, sometimes neighbor links cannot sense each other and they can transmit together. Partial carrier sensing increases the diversity of system states. As a result the two MIS are not as dominant as before. Besides MIS, other states are also likely now. In addition, the transition from one MIS to the other MIS also becomes easier. Assume that with probability $p$ two links can hear each other. When the system state is $\begin{smallmatrix}1&0&1\\0&1&0\end{smallmatrix}$ where link 1, link 4 and link 5 are transmitting, link 2 and link 6 can transmit or actively countdown with probability $(1-p)^2$, and link 3 can be frozen with probability $1-(1-p)^3$. Once one of them begins transmission, with a good chance the system can move to the other MIS $\begin{smallmatrix}0&1&0\\1&0&1\end{smallmatrix}$. Temporal starvation can be alleviated and we conjecture that phase transition as described in [10] will be rather rare in real large wireless networks.

*C. Bandwidth allocation*

Under the 0-1 contention graph model, the links within a clique cannot transmit together, and therefore their aggregate transmission airtime cannot be larger than 1. This has been the basis for many prior investigations on bandwidth/resource allocation in CSMA networks. Due to partial carrier sensing, however, things become more complicated.

Consider a clique of $N$ links. Let $x_i$ be the transmission airtime of link $i$. With full carrier sensing, the constraint is $\sum_{i=1}^{N} x_i \leq 1$ since only one link can transmit each time in a clique. With partial carrier sensing, with probability $p$ two links can sense each other. When link $i$ is transmitting, the other $N-1$ links can transmit or actively countdown with probability $1-p$. When two links are transmitting, the other $N-2$ links can continue to actively countdown with probability $(1-p)^2$, and so on. In a partial carrier-sense network with complex clique formations, the problem of bandwidth allocation will need to be reformulated and re-investigated.

## VI. CONCLUSION

In this paper, we point out that the 0-1 contention graph widely used in the analysis of CSMA networks in research literature is not realistic, and therefore the results and conclusions from such prior studies are suspect. We find that in practice there is a long range of distance where two links partially sense each other (i.e., they sometimes can hear each other, and sometimes cannot). We identify a more accurate *probabilistic* carrier-sensing model in this paper based on real network measurements.

A goal of this paper is to initiate a new research direction so as to take into account the phenomenon of "partial carrier sensing" in future analytical work. To motivate our call for a re-evaluation and a change in direction, we show some examples on how a partial carrier sensing model can lead to conclusions that are different from those obtained under the 0-1 carrier sensing model. For example, an analytical conclusion from the 0-1 model is that link starvations and throughput unfairness can easily arise in many network topologies. We believe that due to partial carrier sensing, starvations and throughput unfairness are actually less common (at least less severe) in real networks.